\documentclass[twoside,dvipdfmx]{LCWS}
\usepackage[utf8]{inputenc}
\usepackage[dvips]{graphicx,epsfig,color}
\usepackage{amssymb,amsmath,array}
\usepackage{cite}
\usepackage{subcaption}

\renewcommand{\theenumi}{\roman{enumi}}

\begin{document}
\title{
One loop effects of natural SUSY in indirect searches for SUSY particles at the ILC
}
\author{M. Jimbo$^1$, T. Kon$^2$, Y. Kouda$^2$, M. Ichikawa$^2$,  
Y. Kurihara$^3$, T. Ishikawa$^3$, K. Kato$^4$, and M. Kuroda$^5$
\vspace{.3cm}\\
\llap{$^1$}  Chiba University of Commerce, Ichikawa, Chiba, Japan
\vspace{.1cm}\\
\llap{$^2$}   Seikei University, Musashino, Tokyo, Japan
\vspace{.1cm}\\
\llap{$^3$}  KEK (Tsukuba), Tsukuba, Ibaraki, Japan
\vspace{.1cm}\\
\llap{$^4$}  Kogakuin University, Shinjuku, Tokyo, Japan
\vspace{.1cm}\\
\llap{$^5$}   Meiji Gakuin University, Yokohama, Kanagawa, Japan\\
}
\maketitle

\begin{abstract}
We have found the possible region of parameters of the minimal supersymmetric standard model (MSSM)
within the bounds from the experimental results of the Higgs mass, the rare decay mode of $b$-quark,
the muon $g-2$, the dark matter abundance,
and the direct searches for the lighter stop (i.e., one of the supersymmetric partners of top quark) at the LHC.
We present numerical results of calculations for the one loop effects of supersymmetric particles in the processes of
$\tau^+ \tau^-$, $b \overline{b}$, $t \overline{t}$, and $Z h$ production at the ILC by using benchmark points
within the possible region of the MSSM parameters.
\end{abstract}

\section{Introduction}
The standard model (SM) of particle physics is considered to be an effective theory
despite the fact that it has succeeded in describing known experimental data available
up to now.  Supersymmetry (SUSY) between bosons and fermions at the unification-energy
scale is believed to be one of the most promising extension of the SM.  Among
the supersymmetric theories, the minimal supersymmetric extension of the SM (MSSM)
is a well studied framework of SUSY because of its compactness.

In the MSSM, however, there are many unfixed parameters.  For limiting the possible
region of the MSSM parameters, a promising approach is so-called ``natural SUSY".
In the framework of the natural SUSY, a light stop with a large A-term and light higgsinos give a solution of fine-tunings
in the MSSM\cite{Kitano:2006}.  We consider that using experimental results is the top priority for limiting the MSSM
parameters, and investigate the possibility of survival of a light stop.

Recently, we have found the possible region of the MSSM parameters\cite{Kouda:2016} within the bounds from
the experimental results of (i) the Higgs mass\cite{HiggsMass:2015}, (ii) the branching ratio,
$\mathrm{Br}(b \rightarrow s \gamma)$\cite{Amhis:2014hma}, (iii) the muon $g-2$\cite{Hoecker:2014},
(iv) the dark matter (DM) abundance\cite{Adam:2015rua,Ade:2015xua}, and
(v) the direct searches for the lighter stop at the LHC\cite{Aad2015,PhysRevD.94.032005,
ATLAS-CONF-2016-076,ATLAS-CONF-2016-050,ATLAS-CONF-2016-077,CMS-PAS-SUS-16-029}.
In the parameter searches, we use {\tt SuSpect2}\cite{suspect2}, {\tt SUSY-HIT}\cite{susyhit}, and
{\tt micrOMEGAs}\cite{micromega} for (i)--(iv).

Moreover, we have studied indirect searches for SUSY particles at the ILC by using benchmark points
within the possible region of the MSSM parameters\cite{Kouda:2016,Kouda:2017,Kon:2017}.  We have calculated
the 1-loop effects of SUSY particles in the processes, $e^+ e^- \rightarrow \tau^+ \tau^-$,
$e^+ e^- \rightarrow b \overline{b}$, $e^+ e^- \rightarrow t \overline{t}$\cite{hollik}, and
$e^+ e^- \rightarrow Z h$\cite{Cao:2014rma,Heinemeyer:2015qbu}
with the aid of {\tt GRACE/SUSY-loop}\cite{gracesusyloop}.

\section{MSSM parameter searches}
Our criterion for searching the MSSM parameters is that the MSSM effects are within the bounds from the
following experimental results:
\begin{enumerate}
\item the Higgs mass, $m_h$,
\item the branching ratio of the rare decay mode of $b$-quark, $\mathrm{Br}(b \rightarrow s \gamma)$,
\item the muon anomalous magnetic moment, $a_{\mu}$,
\item the DM abundance, $\mathnormal{\Omega}_{\mathrm{DM}} h^2$,
\item the lower bounds of the lighter stop mass, $m_{\tilde{t}_1}$ in the direct searches at the LHC.
\end{enumerate}

We have performed systematic multidimensional scans in the entire range of the MSSM parameter
space\cite{Ichikawa:2016}, but we present only the results which are consistent with the five observables above.

\subsection{Higgs mass}
The experimental value of the Higgs mass is measured\cite{HiggsMass:2015} as
\begin{equation}
m_{h}^{\mathrm{exp}} = 125.09 \pm 0.21 \pm 0.11~\mathrm{GeV}\quad . \label{Hm}
\end{equation}

The MSSM contribution to the Higgs mass mainly depends on the masses of stops, $m_{\tilde{t}_1}$,
$m_{\tilde{t}_2}$, the A-term, $A_t$, the higgsino mass parameter, $\mu$, and
$\tan \beta$\cite{Haber:1991PhysRevLett,Okada:1991PTP,Ellis:1991,Haber:1996fp,Carena:2000}.
An alternative parameter, $X_t$ is useful due to the relation among the parameters,
\begin{equation}
X_t \equiv \frac{m_{\tilde{t}_2}^2 - m_{\tilde{t}_1}^2}{m_t} = 2(A_t - \mu \cot \beta)\quad .
\end{equation}

\begin{figure}[htb]
 \centering
 \includegraphics[width=0.7\columnwidth]{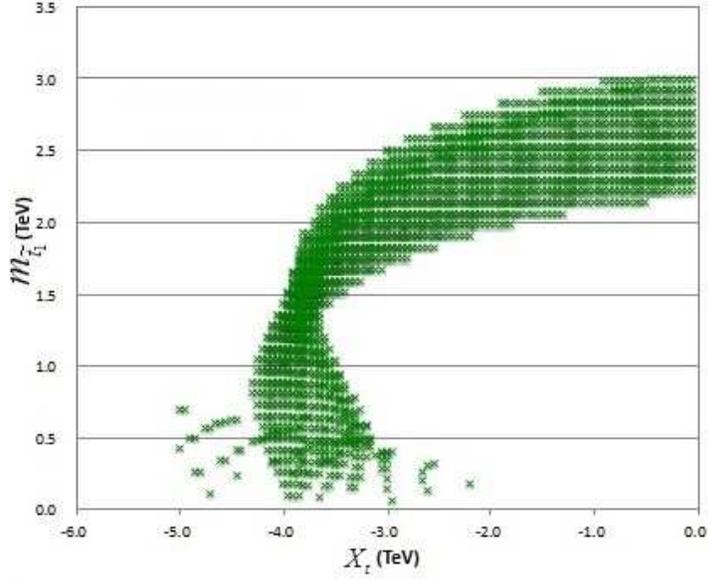}
 \caption{The possible contours on the $X_t$--$m_{\tilde{t}_1}$ plane plotted by green crosses.}
 \label{Fig:xtmst1}
\end{figure}

Figure~\ref{Fig:xtmst1} shows the possible contours on the $X_t$--$m_{\tilde{t}_1}$ plane
which is consistent with \eqref{Hm} for $\tan \beta = 30$ and $\mu = 600~ \mathrm{GeV}$.
The reason for the large value of $\mu$ here is explained in the next subsection.
For the allowed mass of the lighter stop, $m_{\tilde{t}_1}$, there are two possibilities as follows:
\begin{enumerate}
\item the light stop scenario, $m_{\tilde{t}_1} \lesssim 1.5~ \mathrm{TeV}$
with $X_t = -5~ \mathrm{TeV~ to~} -2~ \mathrm{TeV}$,
\item the heavy stop scenario, $m_{\tilde{t}_1} > 1.5~ \mathrm{TeV}$
with $X_t \gtrsim -4~ \mathrm{TeV}$.
\end{enumerate}

\subsection{Rare decay mode of $\boldsymbol{b}$-quark}
\begin{figure}[bht]
 \centering
 \begin{subfigure}{0.4\columnwidth}
  \centering
  \includegraphics[width=\columnwidth]{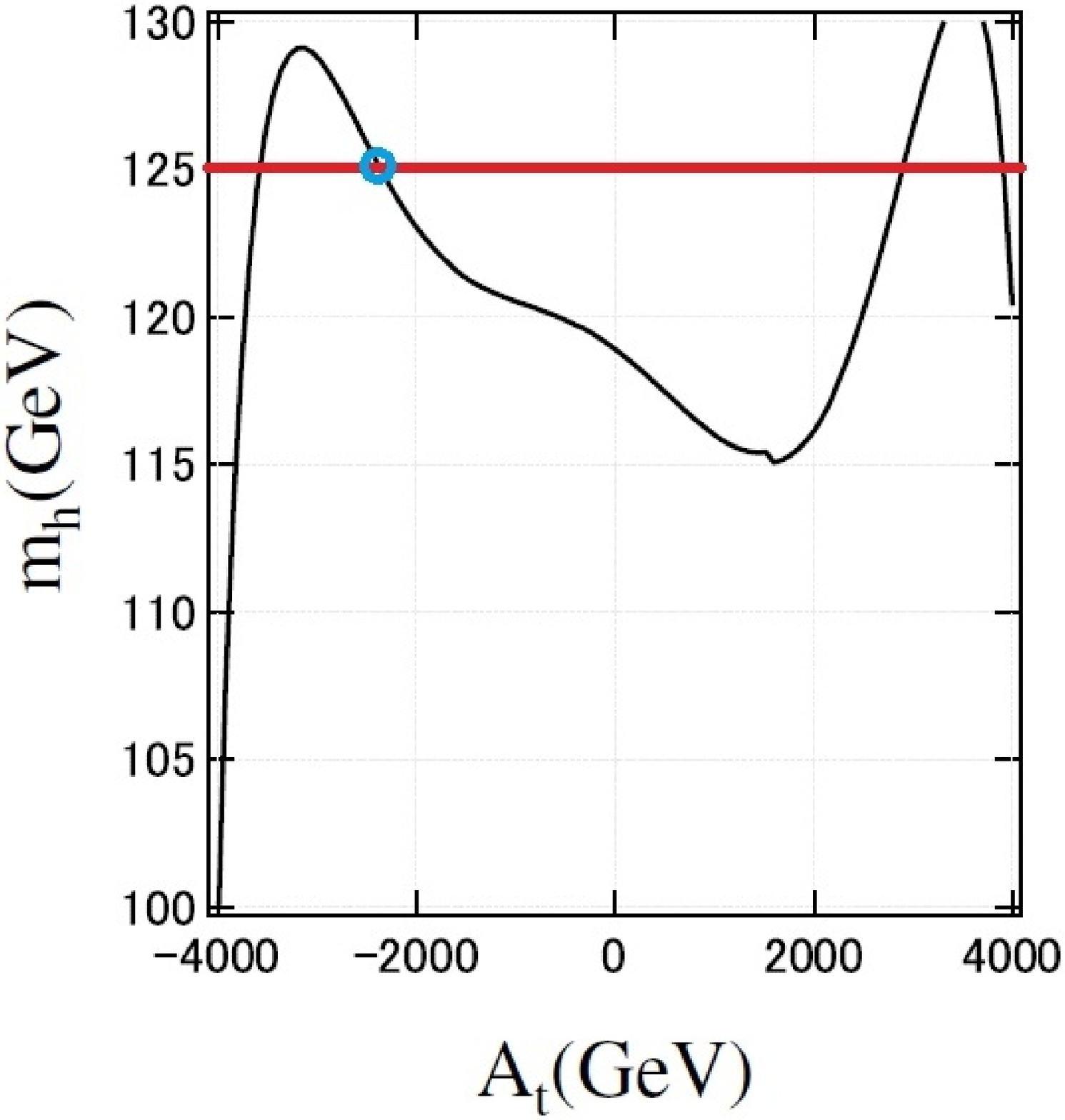}
  \caption{The dependence of the Higgs mass, $m_h$ on $A_t$.  The cyan circle indicates a solution.}
  \label{Fig:Ldetmu}
 \end{subfigure}
 \hspace{2mm}
 \begin{subfigure}{0.4\columnwidth}
  \centering
  \includegraphics[width=\columnwidth]{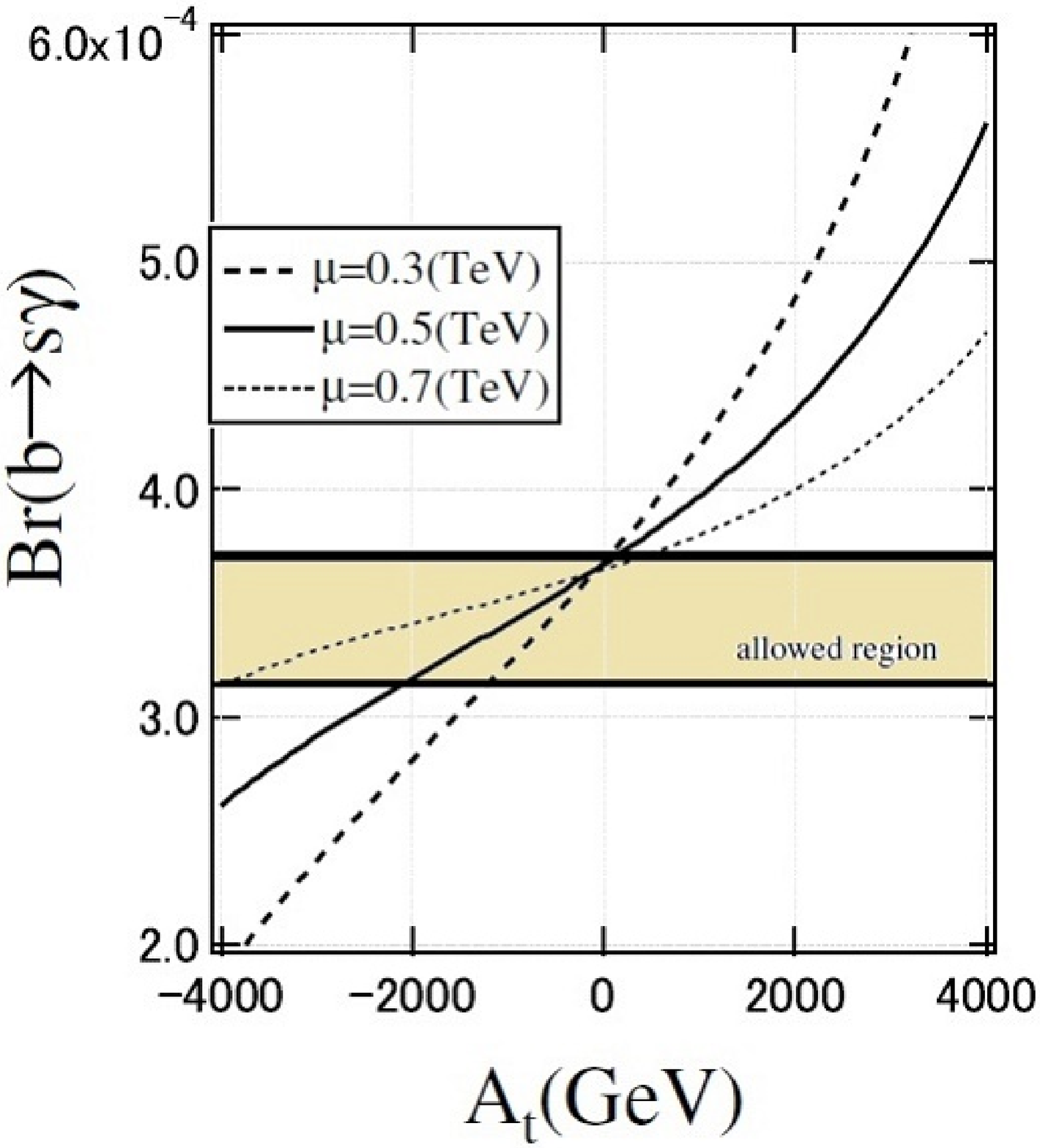}
  \caption{The dependence of the branching ratio, $\mathrm{Br}(b \rightarrow s \gamma)$ on $A_t$.  The allowed
  region is between the two horizontal lines.}
  \label{Fig:Rdetmu}
 \end{subfigure}
 \caption{The constraints on the MSSM parameters, $A_t$ and $\mu$ in the light stop scenario.}
 \label{Fig:detmu}
\end{figure}

At any allowed point in in Figure~\ref{Fig:xtmst1}, the value of $A_t$ is severely restricted by the Higgs mass.
Figure~\ref{Fig:Ldetmu} shows the dependence of the Higgs mass, $m_h$ on $A_t$. Thus, we obtain a solution,
$A_t = -2.4~ \mathrm{TeV}$ for $m_h = 125~ \mathrm{GeV}$. 

The MSSM parameters are also constrained by the experimental value of the branching ratio\cite{Amhis:2014hma},
\begin{equation}
\mathrm{Br}(B \rightarrow X_s \gamma) = (3.43 \pm 0.21 \pm 0.07)\times 10^{-4}\quad .
\end{equation}

Figure~\ref{Fig:Rdetmu} shows the dependence of the branching ratio, $\mathrm{Br}(b \rightarrow s \gamma)$
on $A_t$. Thus, we obtain a constraint, $\mu > 0.5~ \mathrm{TeV}$ for
$A_t = -2.4~ \mathrm{TeV}$.

\clearpage

\subsection{Muon $\boldsymbol{g-2}$}
The muon anomalous magnetic moment, $a_{\mu}\equiv (g_{\mu} - 2)/2$ has been accurately measured,
thus causes the MSSM contribution to be restricted\cite{wang:2015,Endo:2014274}.   The experimental value,
$a_{\mu}^{\mathrm{exp}}$, the SM prediction, $a_{\mu}^{\mathrm{SM}}$, and the difference,
$\Delta a_\mu = a_{\mu}^{\mathrm{exp}} - a_{\mu}^{\mathrm{SM}}$\cite{Hoecker:2014} are
\begin{equation}
 \begin{split}
 a_{\mu}^{\mathrm{exp}} &= (1165920.91 \pm 0.54 \pm 0.33)\times 10^{-9}\quad ,\\
 a_{\mu}^{\mathrm{SM}} &= (1165918.03 \pm 0.01 \pm 0.42 \pm 0.26)\times 10^{-9}\quad ,\\
             \Delta a_\mu &= (2.88 \pm 0.63 \pm 0.49)\times 10^{-9}\quad .
 \end{split}
\end{equation}

The MSSM contribution to $a_{\mu}$ depends on the slepton mass, $m_{\tilde{l}_L}$, $\tan \beta$, $\mu$, the bino mass, $M_1$, and
the wino mass, $M_2$\cite{Cho:2011}.  
\begin{figure}[htb]
 \centering
 \includegraphics[width=0.5\columnwidth]{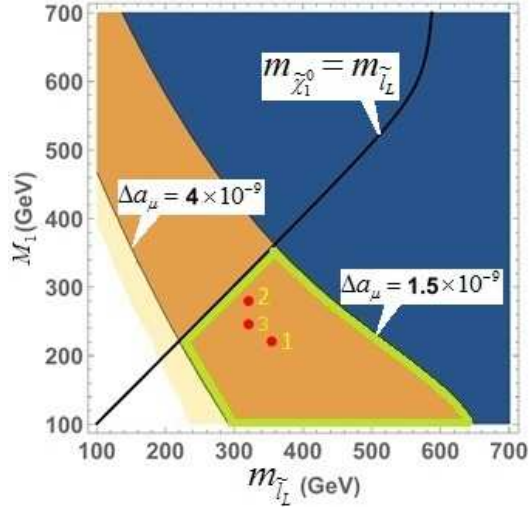}
 \caption{The constraint of the muon $g-2$ for $\tan \beta = 30$,
$\mu = 600~ \mathrm{GeV}$, and $M_2 = 2M_1$. The allowed region is surrounded by the light green lines.
}
 \label{Fig:gminus2}
\end{figure}

Figure~\ref{Fig:gminus2} shows the allowed region on the $m_{\tilde{l}_L}$--$M_1$ plane for $\tan \beta = 30$,
$\mu = 600~ \mathrm{GeV}$, and $M_2 = 2M_1$.  It is remarkable that the bino mass is constrained as $M_1\lesssim 360~ \mathrm{GeV}$, thus the LSP is binolike due to $M_1 < \mu$.
Three red points numbered 1 to 3 are the benchmark points which are defined in
section 2.5\cite{Kouda:2016,Kouda:2017,Kon:2017}.

\subsection{DM abundance}
The lightest SUSY particle (LSP) is a promising candidate for the DM.  We consider the lightest neutralino,
$\tilde{\chi}_1^0$ as the LSP.  The DM abundance can be explained by the LSP-pair annihilation\cite{Ellis:1984453},
the stau--LSP co-annihilation\cite{Ellis:1998367}, and the stop--LSP co-annihilation\cite{Ellis:2014,Ibarra:2015}.
There are three scenarios as in Table~\ref{Scenarios} by combining the light or the heavy stop with the mass
degeneracy of stau--LSP or stop--LSP.  If the masses of stau, stop, and LSP degenerate, both co-annihilation
processes occur.
\begin{table}[htb]
 \caption{Three scenarios for the stau mass and the stop mass.}
 \label{Scenarios}
 \begin{center}
  \begin{tabular}{|c|c|c|} \hline
    & Light stop & Heavy stop \\ \hline
   $m_{\tilde{\tau}_1} \cong m_{\tilde{\chi}_1^0}$ & stau--LSP co-annihilation &  stau--LSP co-annihilation  \\ \hline
   $m_{\tilde{t}_1} \cong m_{\tilde{\chi}_1^0}$ & stop--LSP co-annihilation & --- \\ \hline
  \end{tabular}
 \end{center}
\end{table}

The observed value of the DM abundance\cite{Adam:2015rua,Ade:2015xua},
\begin{equation}
\mathnormal{\Omega}_{\mathrm{DM}}^{\mathrm{exp}} h^2 = 0.1188 \pm 0.0010\quad , \label{EO}
\end{equation}
restricts the MSSM parameters within the narrow region.

\begin{figure}[htb]
 \centering
 \includegraphics[width=0.65\columnwidth]{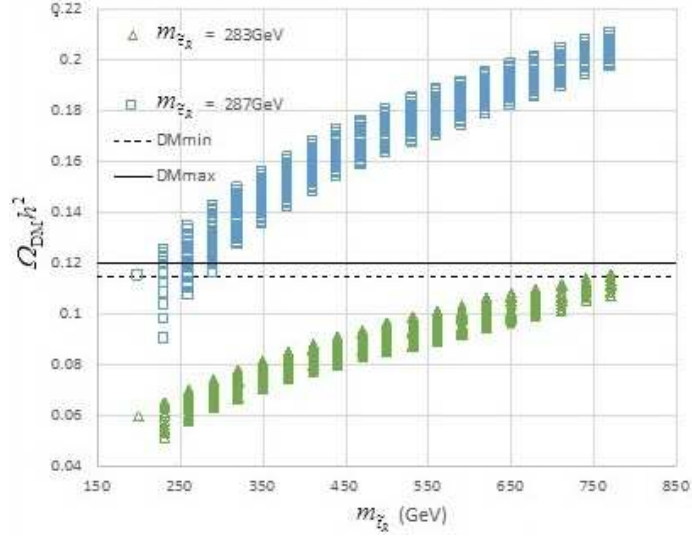}
 \caption{The dependence of the DM abundance on the stop and the stau mass.  The allowed region is between
  the two horizontal lines.}
 \label{Fig:StopDM}
\end{figure}

Figure~\ref{Fig:StopDM} shows the projection of the MSSM parameter scan on the
$m_{\tilde{t}_R}$--$\mathnormal{\Omega}_{\mathrm{DM}} h^2$ plane.  The data are plotted  for two values of
the stau mass, $m_{\tilde{\tau}_R}$.

\begin{figure}[htb]
 \centering
 \includegraphics[width=0.5\columnwidth]{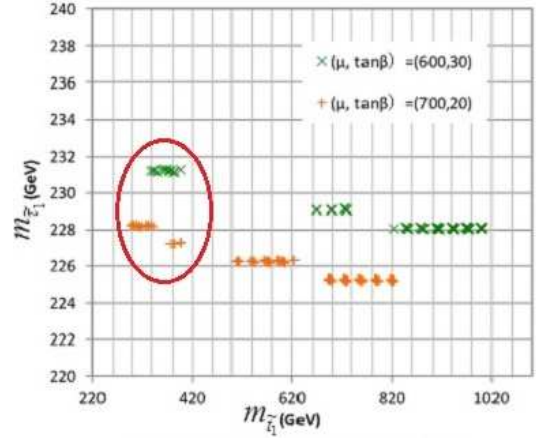}
 \caption{The stop mass and the stau mass consistent with the DM abundance in the light stop scenario
 for $M_1 =  220~ \mathrm{GeV}$.}
 \label{Fig:StopStau}
\end{figure}

Figure~\ref{Fig:StopStau} shows the projection of the MSSM parameter scan on the
$m_{\tilde{t}_1}$--$m_{\tilde{\tau}_1}$ plane which is consistent with \eqref{EO} in the light stop scenario for
$M_1 =  220~ \mathrm{GeV}$.  The data within the red ellipse are consistent with the five observables as mentioned
in the opening paragraph of section 2.  In the allowed region, the lighter stop mass and the lighter stau mass are
restricted as $m_{\tilde{t}_1} = 300~\mathrm{to}~ 400~ \mathrm{GeV}$ and
$m_{\tilde{\tau}_1} = 225~\mathrm{to}~ 235~ \mathrm{GeV}$.  These mass ranges depend on $M_1$.  There,
however, is a room for the heavy stop scenario if $m_{\tilde{\tau}_1} \cong m_{\tilde{\chi}_1^0}$.

\subsection{Direct searches at the LHC}
In the direct searches of SUSY particles at the LHC, there has been no evidence of SUSY particle production
up to now.  We, however, still have a possible region on the MSSM parameters which has not been excluded
by experiments.

\begin{figure}[htb]
 \centering
 \begin{subfigure}{0.32\columnwidth}
  \centering
  \includegraphics[width=\columnwidth]{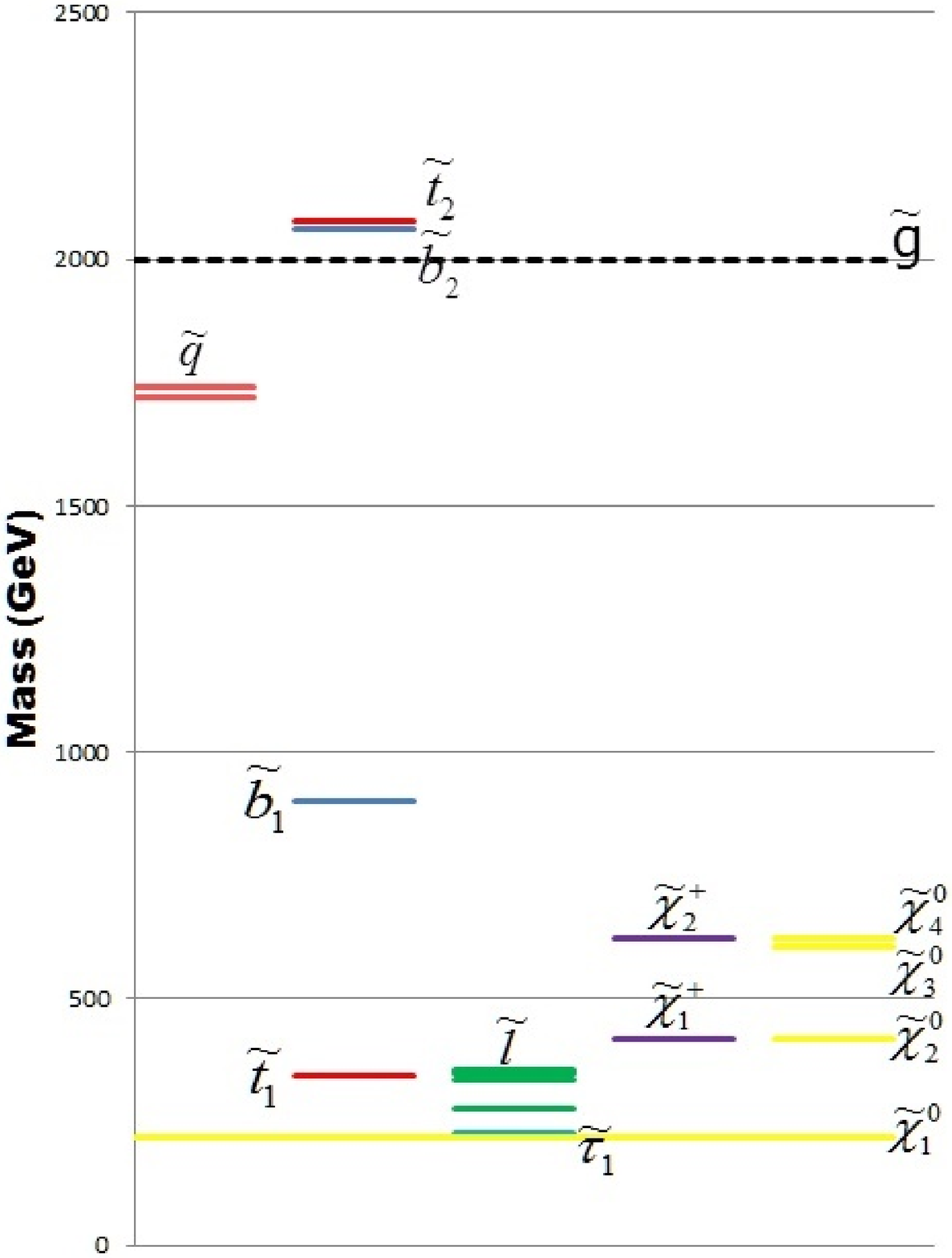}
  \caption{The mass spectrum for set1.}
  \label{Fig:set1}
 \end{subfigure}
 \begin{subfigure}{0.32\columnwidth}
  \centering
  \includegraphics[width=\columnwidth]{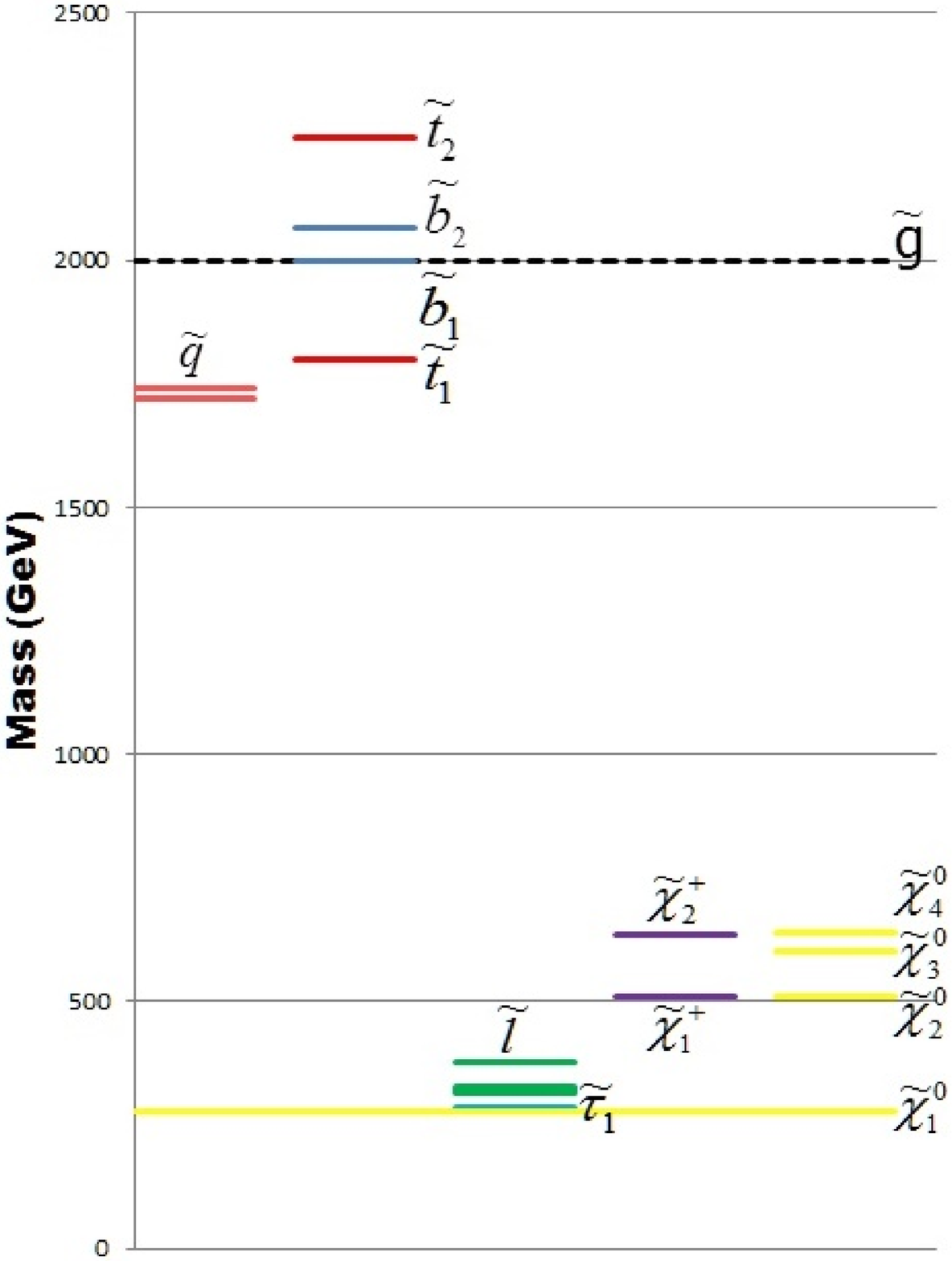}
  \caption{The mass spectrum for set2.}
  \label{Fig:set2}
 \end{subfigure}
 \begin{subfigure}{0.32\columnwidth}
  \centering
  \includegraphics[width=\columnwidth]{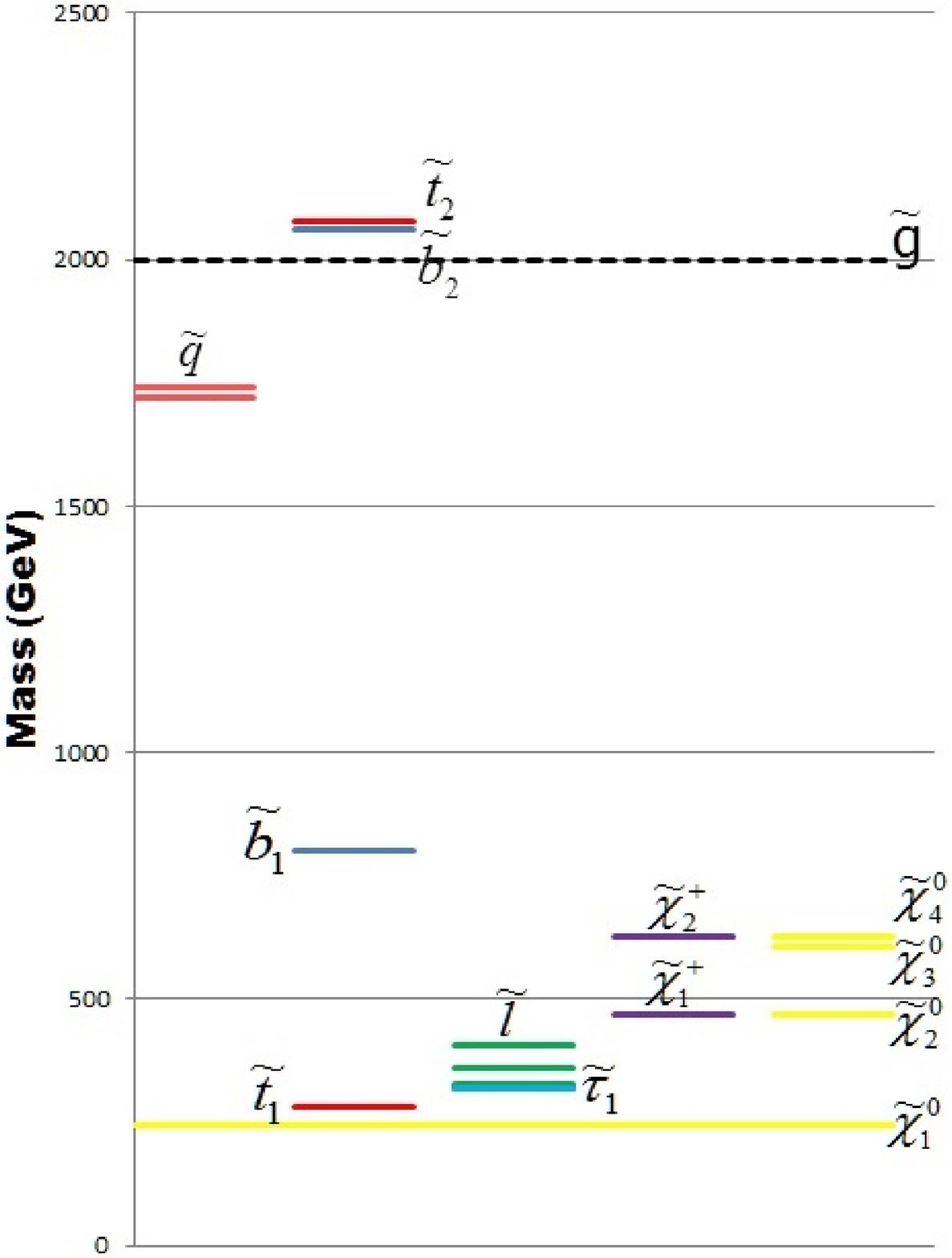}
  \caption{The mass spectrum for set3.}
  \label{Fig:set3}
 \end{subfigure}
 \caption{The mass spectra of the SUSY particles for three parameter sets.}
 \label{Fig:spec}
\end{figure}

Figure~\ref{Fig:spec} shows the mass spectra for the benchmark points which we take from the possible region of
the MSSM parameters corresponding to three scenarios in Table~\ref{Scenarios}.  The MSSM parameter sets
named set1 and set3 are for the light stop scenario, and that named set2 is for the heavy stop scenario.

\begin{figure}[hbt]
 \centering
 \begin{subfigure}{0.45\columnwidth}
  \centering
  \includegraphics[width=\columnwidth]{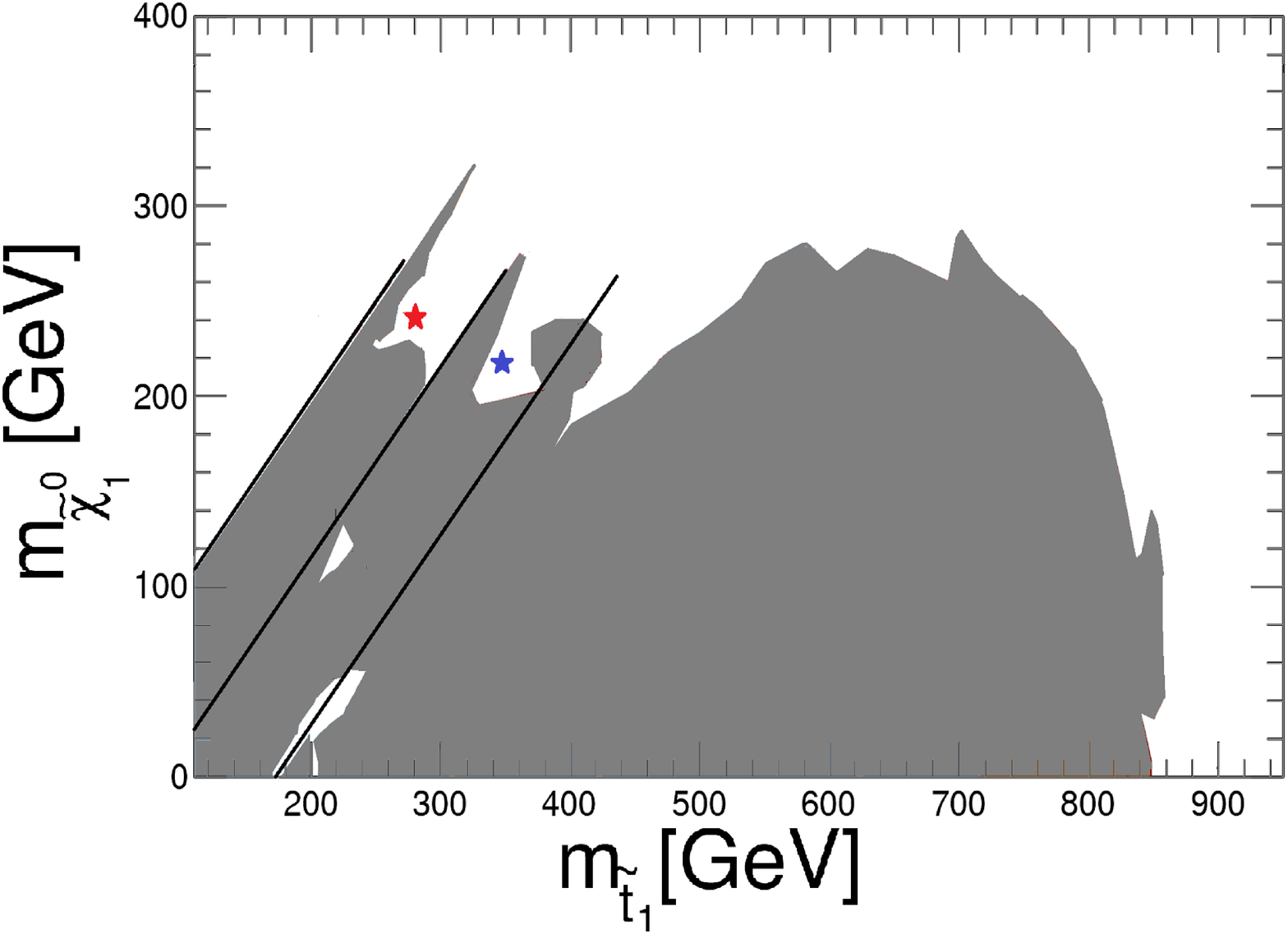}
  \caption{The lighter stop searches in the ATLAS experiments\cite{Aad2015,PhysRevD.94.032005,
ATLAS-CONF-2016-076,ATLAS-CONF-2016-050,ATLAS-CONF-2016-077}.}
  \label{Fig:ATLAS}
 \end{subfigure}
 \hspace{2mm}
 \begin{subfigure}{0.45\columnwidth}
  \centering
  \includegraphics[width=\columnwidth]{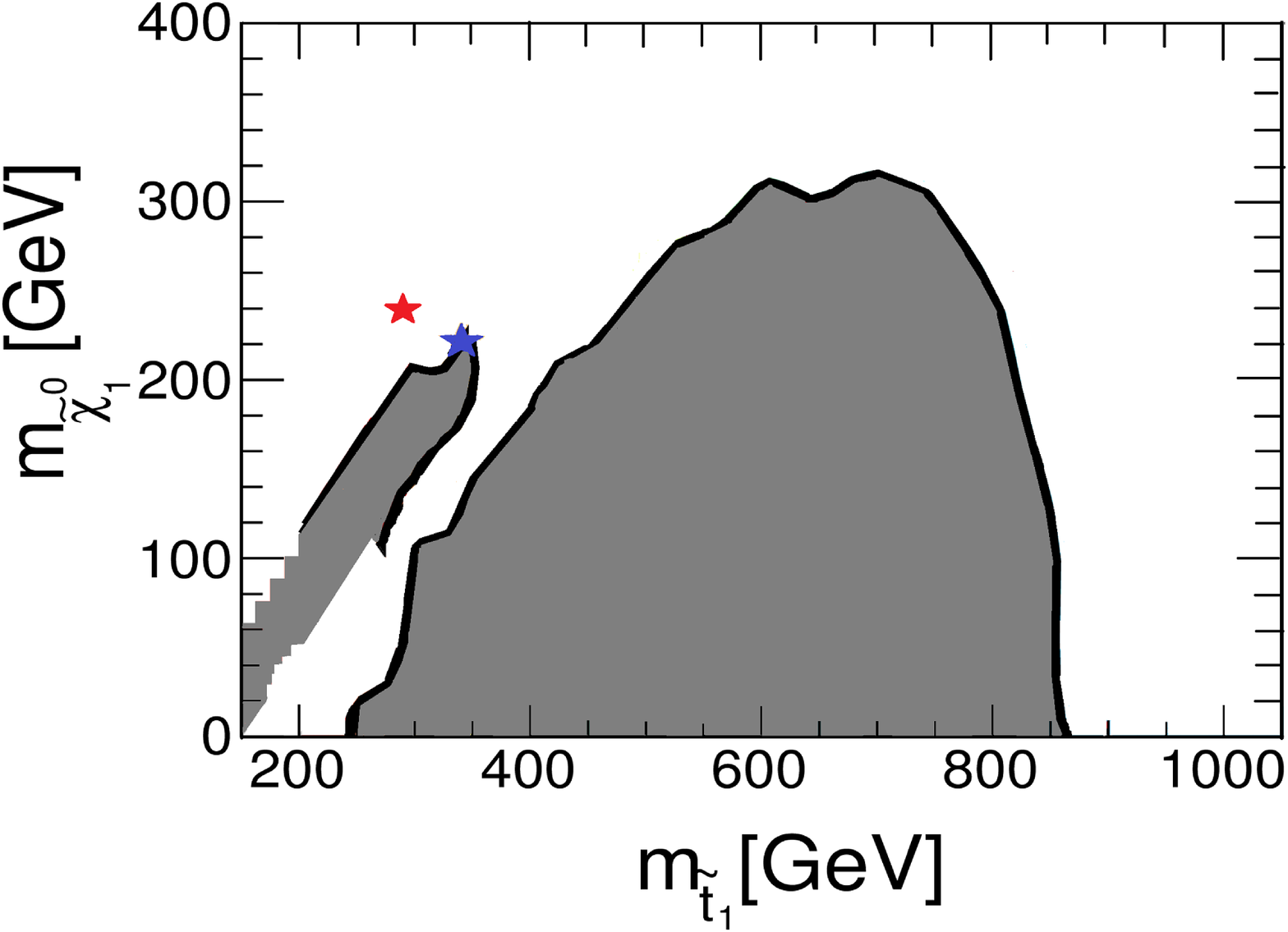}
  \caption{The lighter stop searches in the CMS experiments\cite{CMS-PAS-SUS-16-029}.}
  \label{Fig:CMS}
 \end{subfigure}
 \caption{The lighter stop searches at the LHC.  The gray regions are excluded by experiments.  The blue star and
the red one indicate the benchmark points, set1 and set3, respectively.  The benchmark point, set2 is out
of the ranges of the graphs because of the heavy stop mass, $M_{\tilde{t}_1} = 1799~ \mathrm{GeV}.$
}
 \label{Fig:LHC}
\end{figure}

Figure~\ref{Fig:LHC} shows the mass bounds of the lighter stop, $\tilde{t}_1$ from the direct searches at the
LHC\cite{Aad2015,PhysRevD.94.032005,ATLAS-CONF-2016-076,ATLAS-CONF-2016-050,ATLAS-CONF-2016-077,
CMS-PAS-SUS-16-029}.  The two benchmark points plotted in the figure are located outside of the excluded regions.
Thus we find that two kinds of the light stop scenarios and the heavy stop scenario still survive.

\section{Numerical results}
For indirect searches of SUSY particles, SUSY effects are revealed through 1-loop corrections.  There, however,
exist the SM contributions to 1-loop corrections, thus the following quantities are useful\cite{hollik}.
\begin{align}
 \delta_{\mathrm{susy}}^{\mathrm{EW}} & \equiv \frac{d \sigma_{\mathrm{1-loop}}^{\mathrm{MSSM, EW}}
                              - d \sigma_{\mathrm{1-loop}}^{\mathrm{SM, EW}}}{d \sigma_{\mathrm{tree}}}\quad ,\\
 \delta_{\mathrm{susy}}^{\mathrm{QCD}} & \equiv \frac{d\sigma_{\mathrm{1-loop}}^{\mathrm{MSSM, QCD}}
                               - d \sigma_{\mathrm{1-loop}}^{\mathrm{SM, QCD}}}{d \sigma_{\mathrm{tree}}}\quad ,\\
 \delta_{\mathrm{susy}} & \equiv \delta_{\mathrm{susy}}^{\mathrm{EW}} + \delta_{\mathrm{susy}}^{\mathrm{QCD}}
                               \quad .
\end{align}
\begin{figure}[hbt]
 \centering
 \includegraphics[width=0.38\columnwidth]{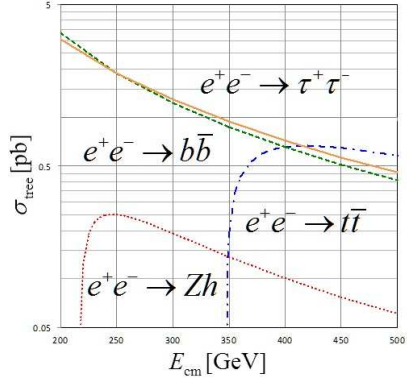}
 \caption{The dependence of the tree level cross sections on $E_{\mathrm{cm}}$.}
 \label{Fig:tree}
\end{figure}

For the indirect searches of SUSY particles, we consider the following processes:
\begin{align*}
(\mathrm{i})~ & e^+ + e^- \rightarrow \tau^+ + \tau^- \qquad& \mathrm{at}~ E_{\mathrm{cm}} = 250~ \mathrm{GeV}
\quad , & \hspace{145pt}~ \\
(\mathrm{ii})~ & e^+ + e^- \rightarrow b + \overline{b} \qquad & \mathrm{at}~ E_{\mathrm{cm}} = 250~ \mathrm{GeV}
\quad , & \hspace{145pt}~ \\
(\mathrm{iii})~ & e^+ + e^- \rightarrow t + \overline{t} \qquad & \mathrm{at}~ E_{\mathrm{cm}} = 500~ \mathrm{GeV}
\quad , & \hspace{145pt}~ \\
(\mathrm{iv})~ & e^+ + e^- \rightarrow Z + h \qquad & \mathrm{at}~ E_{\mathrm{cm}} = 250~ \mathrm{GeV}\quad ,
 & \hspace{145pt}~
\end{align*}
where the center-of-mass energy, $E_{\mathrm{cm}}$ is determined by the tree level cross sections as in
Figure~\ref{Fig:tree} and by the technical design of the ILC\cite{ilcd}.

\begin{figure}[hbt]
 \centering
 \begin{subfigure}{0.45\columnwidth}
  \centering
  \includegraphics[width=\columnwidth]{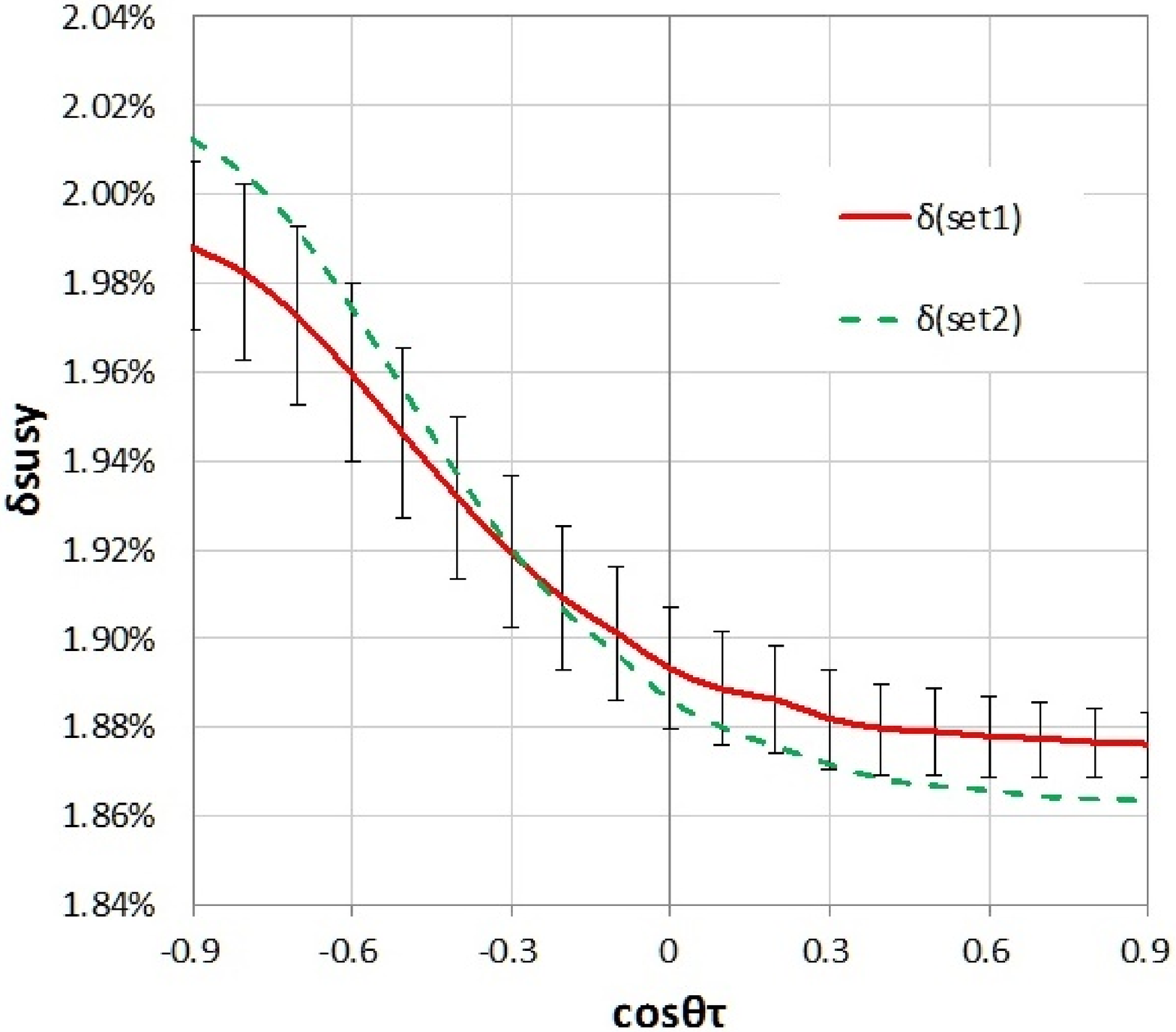}
  \caption{The angular distribution of $\delta_{\mathrm{susy}}$ in the $\tau^+ \tau^-$-pair production process.}
  \label{Fig:TAUtau}
 \end{subfigure}
 \hspace{2mm}
 \begin{subfigure}{0.45\columnwidth}
  \centering
  \includegraphics[width=\columnwidth]{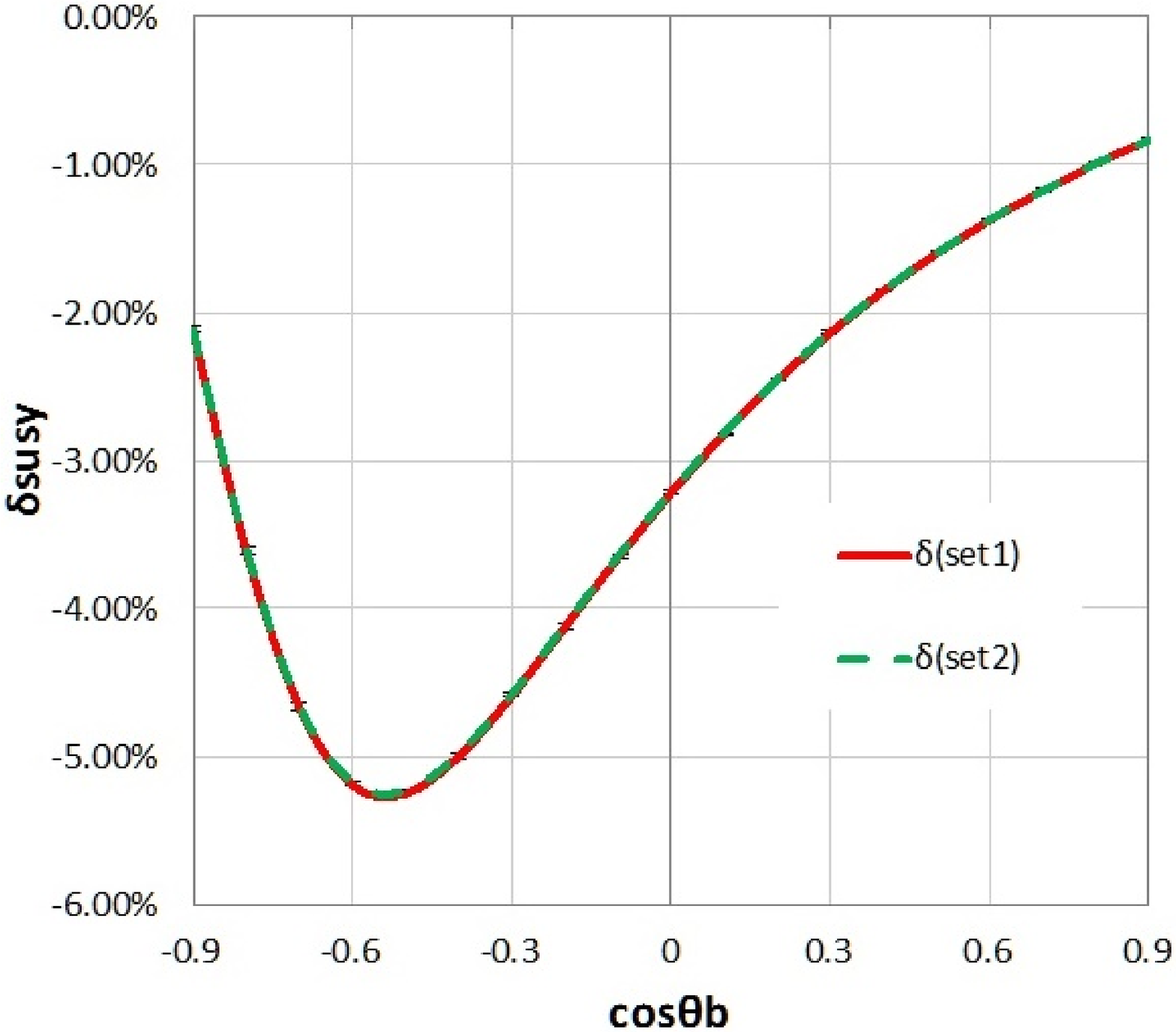}
  \caption{The angular distribution of $\delta_{\mathrm{susy}}$ in the $b \overline{b}$-pair production process.}
  \label{Fig:bB}
 \end{subfigure}
 \caption{The indirect searches of SUSY particles for the parameter sets, set1 and set2 at
  $E_{\mathrm{cm}} = 250~ \mathrm{GeV}$.  In (a), statistical errors are shown only for the results of set1
where we assumed that the integrated luminosity is  $250 \mathrm{fb^{-1}}$.}
 \label{Fig:Ecm250}
\end{figure}

Figure~\ref{Fig:TAUtau} and Figure~\ref{Fig:bB} show the angular distributions of $\delta_{\mathrm{susy}}$
in the $\tau^+ \tau^-$-pair production process and the $b \overline{b}$-pair production one, respectively.
It is remarkable that SUSY effects can be revealed as excess of the SM estimation at the 1-loop level,
because the value of $\delta_{\mathrm{susy}}$ is larger than the expected error bars.  For the integrated
luminosity, $\int L dt = 250 \mathrm{fb^{-1}}$, however, the parameter set of the light stop scenario, set1 cannot be
discriminated from that of the heavy stop scenario, set2 because of the mass relations,
$m_{\tilde{\tau}_1} \cong m_{\tilde{\chi}_1^0}$ and $m_{\tilde{b}_1} \gg E_{\mathrm{cm}}$ in both parameter sets.

\begin{figure}[htb]
 \centering
 \includegraphics[width=0.43\columnwidth]{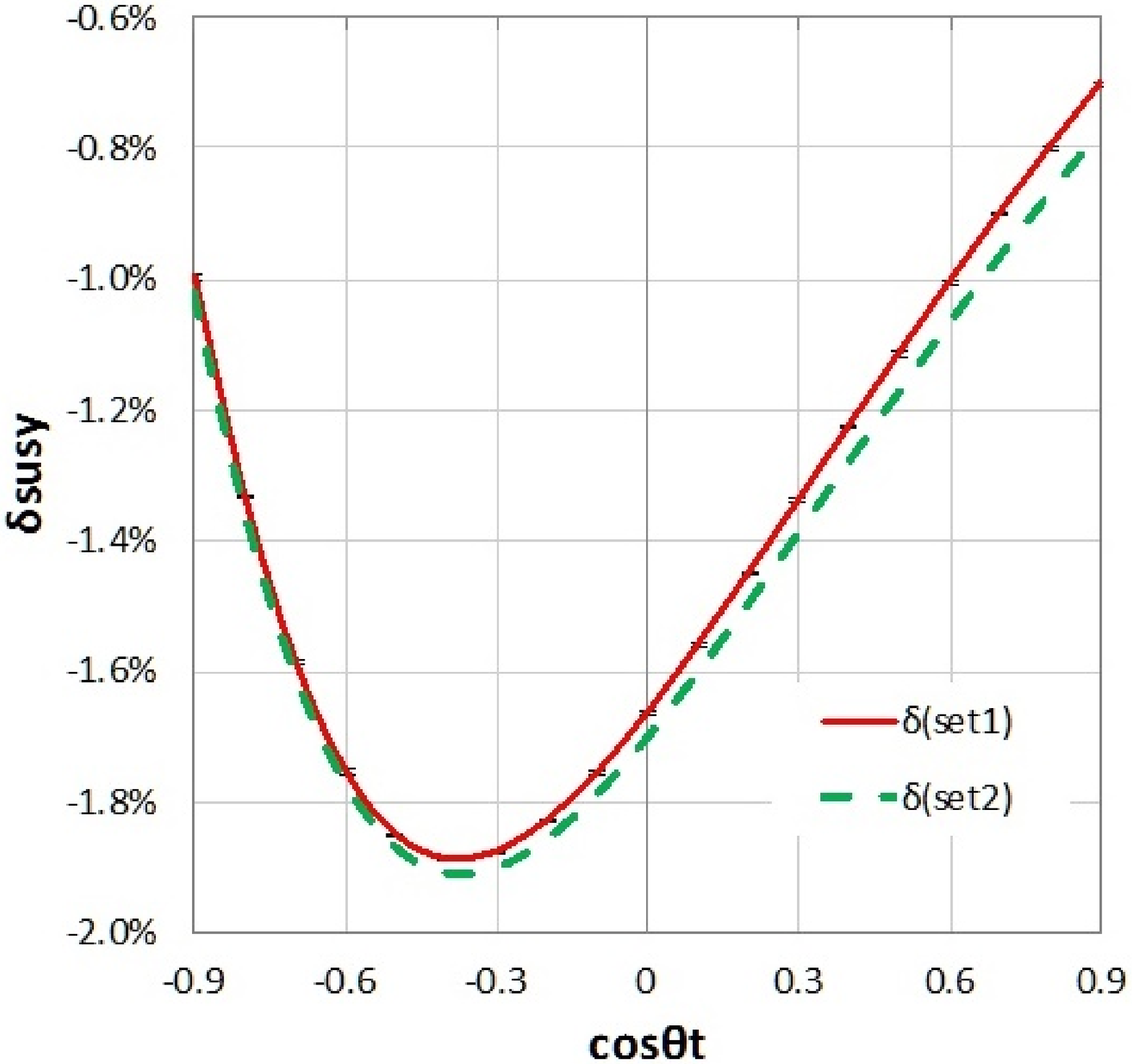}
 \caption{The angular distribution of $\delta_{\mathrm{susy}}$ in the $t \overline{t}$-pair production process for
  the parameter sets, set1 and set2 at $E_{\mathrm{cm}} = 500~ \mathrm{GeV}$.  Statistical errors are shown only
for the results of set1 where we assumed that the integrated luminosity is  $500 \mathrm{fb^{-1}}$,
but are negligibly small.}
 \label{Fig:tT500}
\end{figure}

Figure~\ref{Fig:tT500} shows the angular distribution of $\delta_{\mathrm{susy}}$ in the $t \overline{t}$-pair production
process.  There is small difference between the $\delta_{\mathrm{susy}}$'s of the two parameter sets in the forward direction of top quark.

\clearpage

\begin{figure}[hbt]
 \centering
 \includegraphics[width=0.43\columnwidth]{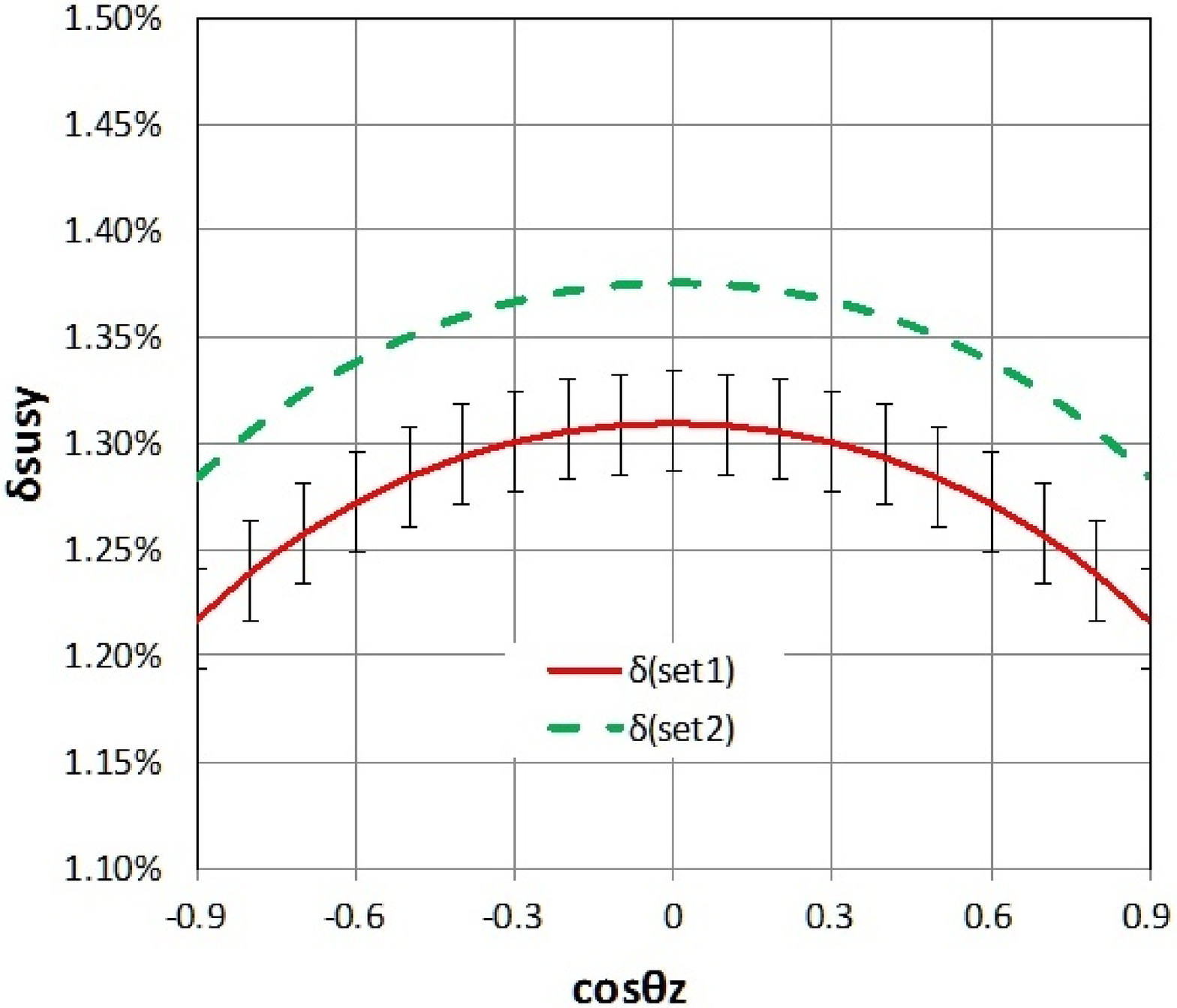}
 \caption{The angular distribution of $\delta_{\mathrm{susy}}$ in the $Z h$ production process for the parameter sets,
  set1 and set2 at $E_{\mathrm{cm}} = 250~ \mathrm{GeV}$.  Statistical errors are shown only
for the results of set1 where we assumed that the integrated luminosity is  $250 \mathrm{fb^{-1}}$.}
 \label{Fig:Zh250}
\end{figure}

\clearpage

Figure~\ref{Fig:Zh250} shows the angular distribution of $\delta_{\mathrm{susy}}$ in the $Z h$ production process.
There is difference between the $\delta_{\mathrm{susy}}$'s of the two parameter sets, which is larger than the range
of error bars.  It is remarkable that the 1-loop corrections of this process are sensitive to the lighter stop mass, $m_{\tilde{t}_1}$.

\section{Summary}
We have obtained possible MSSM parameter sets which are consistent with the bounds from the experimental
results of the Higgs mass, the rare decay mode of $b$-quark, the muon $g-2$, the dark matter abundance, and the direct searches for the lighter stop at the LHC.
For the parameter sets, the virtual effects of MSSM in the indirect search processes are estimated at 1\%$\ \sim 5$\%
in the early stage of the ILC, which are larger than the ranges of error bars.  For the purpose of discrimination of the light and heavy stop scenarios, $Z h$ is the most promising process to investigate.


\end{document}